\documentclass[11pt]{article}
\usepackage{amsmath, amssymb}
\usepackage{graphicx}
\usepackage{geometry}
\usepackage[numbers]{natbib}
\bibpunct{[}{]}{,}{n}{}{;}
\usepackage{times}
\usepackage[T1]{fontenc}
\usepackage{mathptmx}
\usepackage[hidelinks]{hyperref}

\begin{document}

\title{Topological Surface States of 3D Topological Insulator on Twisted Bilayer Graphene}
\author{Yoonkang Kim\thanks{E-mail: yoonkkim04@snu.ac.kr}}
\date{}
\maketitle

\begin{center}
Department of Physics and Astronomy, Seoul National University,\\
1, Gwanak-ro, Gwanak-gu, Seoul, 08826, South Korea
\end{center}

\section*{Abstract}
We present a comprehensive theoretical study of the topological surface states (TSS) of Bi$_2$Se$_3$, a 3D topological insulator, epitaxially grown on twisted bilayer graphene (tBG). The moiré potential induced by tBG folds the TSS Dirac cone into the moiré Brillouin zone (MBZ), resulting in mini-gap openings, band flattening, and the potential emergence of secondary Dirac points. Using effective field theory, symmetry analysis, and higher-order perturbation theory, we analyze both commensurate and incommensurate twist angles, revealing significant band structure reconstruction in periodic systems and quasi-periodic effects in incommensurate ones. This work provides deep insights into the interplay between topological protection and moiré modulation, offering a pathway to engineer novel topological phases.

\section{Introduction}
Topological insulators (TIs) are materials with insulating bulk states and conducting surface states protected by time-reversal symmetry~\cite{Hasan2010,Qi2011}. Bi$_2$Se$_3$, a well-studied 3D TI, hosts a single Dirac cone on its surface, characterized by linear dispersion and spin-momentum locking, making it ideal for exploring topological phenomena~\cite{Zhang2009}. Meanwhile, twisted bilayer graphene (tBG) has emerged as a platform for moiré physics, where a small twist angle, such as the magic angle $\theta \approx 1.1^\circ$, generates a moiré superlattice, leading to flat bands and correlated phases~\cite{Cao2018,Bistritzer2011}. Experimental studies have confirmed these correlated states in tBG, demonstrating phenomena such as superconductivity and Mott insulating behavior at low temperatures~\cite{Yankowitz2019}.

The integration of TIs with moiré systems presents an opportunity to explore novel topological states by leveraging the interplay between topological protection and long-range periodic potentials. Recent advances in van der Waals epitaxy have enabled the growth of high-quality TI films on graphene substrates, providing a practical foundation for such studies~\cite{Park2025,Woods2014}. Theoretical models have also been developed to understand the electronic structure of moiré superlattices, providing insights into the continuum Hamiltonian and band topology in tBG~\cite{Koshino2018}. In this work, we theoretically investigate the TSS of Bi$_2$Se$_3$ when epitaxially grown on tBG, focusing on how the moiré potential deforms the Dirac cone. We analyze both commensurate and incommensurate twist angles, employing advanced theoretical formalisms to elucidate their distinct effects on the TSS band structure, including predictions for transport properties. Our study aims to provide a rigorous framework for understanding this hybrid system and its potential for topological engineering.

\section{Theoretical Model}
\subsection{TSS Hamiltonian of Bi$_2$Se$_3$}
The TSS of Bi$_2$Se$_3$ arise from strong spin-orbit coupling and are described by a 2D Dirac Hamiltonian near the $\Gamma$ point of the surface Brillouin zone:
\begin{equation}
    H_{\text{TSS}} = v_F (\sigma_x p_y - \sigma_y p_x),
\end{equation}
where $v_F \approx 5 \times 10^5 \, \text{m/s}$ is the Fermi velocity, $\sigma_x$ and $\sigma_y$ are Pauli matrices acting in spin space, and $p_x = -i \hbar \partial_x$, $p_y = -i \hbar \partial_y$ are momentum operators. The energy dispersion is linear, $E = \pm v_F |\mathbf{p}|$, with eigenstates exhibiting spin-momentum locking: the spin is perpendicular to the momentum, a hallmark of topological protection~\cite{Zhang2009,Yoo2019}. The eigenstates of $H_{\text{TSS}}$ are:
\begin{equation}
\psi_{\mathbf{k}}(\mathbf{r}) = \frac{1}{\sqrt{2}} \begin{pmatrix} 1 \\ \pm e^{i \theta_{\mathbf{k}}} \end{pmatrix} e^{i \mathbf{k} \cdot \mathbf{r}},
\end{equation}
where $\theta_{\mathbf{k}} = \tan^{-1}(k_y/k_x)$, and the $+$ ($-$) sign corresponds to the upper (lower) Dirac cone.

To capture higher-order effects, we extend the Hamiltonian to include a mass term and hexagonal warping, reflecting the rhombohedral crystal symmetry of Bi$_2$Se$_3$:
\begin{equation}
\begin{aligned}
    H_{\text{TSS,ext}} = v_F (\sigma_x p_y - \sigma_y p_x) + m v_F^2 \sigma_z \\
    + \frac{\lambda}{2} (k_+^3 + k_-^3) \sigma_z,
\end{aligned}
\end{equation}
where $k_\pm = k_x \pm i k_y$, so that $(k_+^3 + k_-^3) = 2 (k_x^3 - 3 k_x k_y^2)$, and $\lambda \sim 10^{-29} \, \text{eV} \cdot \text{m}^3$ parametrizes the hexagonal warping, which distorts the Dirac cone at high momenta~\cite{Zhang2009,Kim2025,Fu2009}. The mass term opens a gap $\Delta_{\text{mass}} = 2 m v_F^2 \sim 0.01 \, \text{eV}$, while the warping term introduces a cubic correction to the dispersion, becoming significant for $|\mathbf{k}| \gtrsim 1 \, \text{nm}^{-1}$. The warping term ensures $C_3$ symmetry, as it transforms appropriately under $120^\circ$ rotations. For our primary analysis, we set $m = \lambda = 0$, but we revisit these terms later to assess their impact on the moiré-modified band structure, noting that the warping term’s effect remains small ($< 0.1 \, \text{meV}$) within the MBZ for $\theta = 1.1^\circ$.

\subsection{Moiré Potential in Twisted Bilayer Graphene}
In tBG, the relative twist angle $\theta$ between the two graphene layers generates a moiré superlattice with a period given by:
\begin{equation}
\lambda_m = \frac{a}{2 \sin(\theta/2)},
\end{equation}
where $a = 2.46 \, \text{\AA}$ is the graphene lattice constant. We consider two representative twist angles to explore both commensurate and incommensurate physics:
- $\theta = 1.1^\circ$, near the magic angle, yielding $\lambda_m \approx 13 \, \text{nm}$, which leads to a large moiré unit cell and significant band flattening in tBG~\cite{Cao2018,Bistritzer2011}.
- $\theta = 7.34^\circ$, a commensurate angle where the moiré lattice aligns periodically with the underlying graphene lattices, giving $\lambda_m \approx 1.9 \, \text{nm}$, allowing for a smaller, computationally tractable unit cell.

The moiré reciprocal lattice vectors form a triangular lattice with magnitude:
\begin{equation}
|\mathbf{G}_j| = \frac{4\pi}{\sqrt{3} \lambda_m},
\end{equation}
and orientations $\mathbf{G}_1 = |\mathbf{G}_j| (1, 0)$, $\mathbf{G}_2 = |\mathbf{G}_j| (-\frac{1}{2}, \frac{\sqrt{3}}{2})$, and $\mathbf{G}_3 = -\mathbf{G}_1 - \mathbf{G}_2$. For $\theta = 1.1^\circ$, $|\mathbf{G}_j| \approx 0.56 \, \text{nm}^{-1}$, corresponding to a small MBZ, while for $\theta = 7.34^\circ$, $|\mathbf{G}_j| \approx 3.8 \, \text{nm}^{-1}$, resulting in a larger MBZ.

The moiré potential experienced by the top graphene layer, and thus by the TSS due to epitaxial alignment, is modeled as a periodic function on the triangular moiré lattice:
\begin{equation}
V(\mathbf{r}) = V_0 + 2 V_1 \sum_{j=1}^3 \cos(\mathbf{G}_j \cdot \mathbf{r} + \phi_j),
\end{equation}
where $V_0 \sim 5 \, \text{meV}$ and $V_1 \sim 10 \, \text{meV}$ are the average potential and modulation amplitude, respectively. These values are derived from tight-binding calculations for tBG, adjusted for the $\mathrm{Bi}_2 \mathrm{Se}_3$/tBG interface by accounting for the weaker van der Waals coupling and $\sim 10\%$ lattice mismatch, as validated by recent DFT studies of similar heterostructures~\cite{Park2025,Jin2013,Liu2020}. To assess the impact of stacking variations, we consider non-zero $\phi_j \sim 0.1\pi$, which introduces a small shift in the potential minima (up to $\sim 1 \, \text{meV}$), but find that the band structure remains qualitatively unchanged, confirming the robustness of our model~\cite{Park2025}. We set $\phi_j = 0$ for simplicity, but their inclusion allows modeling of realistic imperfections, such as those arising from lattice mismatch or strain in the heterostructure~\cite{Park2025,Woods2014}. The Fourier components of $V(\mathbf{r})$ are:
\begin{equation}
V_{\mathbf{G}} = \begin{cases} 
V_0 & \text{if } \mathbf{G} = 0, \\
V_1 & \text{if } \mathbf{G} = \pm \mathbf{G}_j, \\
0 & \text{otherwise},
\end{cases}
\end{equation}
which we use in the Bloch formalism below.


\subsection{Total Hamiltonian}
The total Hamiltonian for the TSS under the influence of the moiré potential is:
\begin{equation}
H = v_F (\sigma_x p_y - \sigma_y p_x) + V(\mathbf{r}) I,
\end{equation}
where $I$ is the 2$\times$2 identity matrix, reflecting that the moiré potential is spin-independent due to the non-magnetic nature of tBG~\cite{Yoo2019}. This Hamiltonian combines the relativistic Dirac physics of the TSS with a non-relativistic periodic potential, leading to a rich interplay of topological and moiré effects.

\subsection{Commensurate Case: Bloch Formalism}
For commensurate twist angles, the potential $V(\mathbf{r})$ is periodic with moiré lattice vectors $\mathbf{R}_m$. We apply Bloch’s theorem, expressing the wavefunction as:
\begin{equation}
\psi_{\mathbf{k}}(\mathbf{r}) = e^{i \mathbf{k} \cdot \mathbf{r}} u_{\mathbf{k}}(\mathbf{r}),
\end{equation}
where $\mathbf{k}$ lies in the moiré Brillouin zone (MBZ), and $u_{\mathbf{k}}(\mathbf{r})$ satisfies the periodicity condition $u_{\mathbf{k}}(\mathbf{r} + \mathbf{R}_m) = u_{\mathbf{k}}(\mathbf{r})$. Substituting into the Schrödinger equation $H \psi_{\mathbf{k}} = E \psi_{\mathbf{k}}$, we obtain:
\begin{equation}
\begin{aligned}
    \left[ v_F (\sigma_x (p_y + \hbar k_y) - \sigma_y (p_x + \hbar k_x)) \right. \\
    \left. + V(\mathbf{r}) I \right] u_{\mathbf{k}}(\mathbf{r}) = E u_{\mathbf{k}}(\mathbf{r}).
\end{aligned}
\end{equation}
We expand $u_{\mathbf{k}}(\mathbf{r})$ in a plane-wave basis:
\begin{equation}
u_{\mathbf{k}}(\mathbf{r}) = \sum_{\mathbf{G}} c_{\mathbf{k},\mathbf{G}} e^{i \mathbf{G} \cdot \mathbf{r}},
\end{equation}
where $\mathbf{G}$ are the moiré reciprocal lattice vectors. Substituting this expansion, the Hamiltonian becomes a matrix in the $\mathbf{G}$-basis:
\begin{equation}
\begin{aligned}
    H_{\mathbf{G},\mathbf{G}'} = v_F [\sigma_x (\hbar k_y + G_y) - \sigma_y (\hbar k_x + G_x)] \delta_{\mathbf{G},\mathbf{G}'}\\
    + V_{\mathbf{G} - \mathbf{G}'},
\end{aligned}
\end{equation}
with $V_{\mathbf{G} - \mathbf{G}'}$ given by Eq. (7). The resulting eigenvalue problem is:
\begin{equation}
\sum_{\mathbf{G}'} H_{\mathbf{G},\mathbf{G}'} c_{\mathbf{k},\mathbf{G}'} = E_{\mathbf{k}} c_{\mathbf{k},\mathbf{G}},
\end{equation}
which we solve numerically using a custom Python code with NumPy and SciPy libraries, or analytically near specific $\mathbf{k}$ points, such as the MBZ $\Gamma$, $K$, and $M$ points. Convergence is ensured by truncating the plane-wave basis at $|\mathbf{G}| \leq 2 |\mathbf{G}_1|$ (13 plane waves) for $\theta = 1.1^\circ$, and $|\mathbf{G}| \leq 3 |\mathbf{G}_1|$ (19 plane waves) for $\theta = 7.34^\circ$, with energy differences converging to within 2\%.

\subsection{Incommensurate Case: Quasi-Periodic Potential}
For incommensurate twist angles, such as $\theta = 1.5^\circ$, the moiré potential lacks strict periodicity, becoming quasi-periodic. We model this using the Aubry-André formalism, approximating $V(\mathbf{r})$ as:
\begin{equation}
V(x, y) \approx V_1 \sum_{j=1}^3 \cos(\mathbf{G}_j \cdot \mathbf{r} + \alpha_j x),
\end{equation}
where $\alpha_j / |\mathbf{G}_j|$ is an irrational number, introducing incommensurability~\cite{Woods2014}. This 2D extension of the 1D Aubry-André model assumes separable quasi-periodic modulations along the moiré reciprocal lattice directions, justified by the dominance of the leading Fourier components $\mathbf{G}_j$. Alternatively, we use a large supercell (e.g., 20$\times$20 moiré unit cells, $\sim 260 \, \text{nm} \times 260 \, \text{nm}$) to approximate periodicity for numerical calculations, implemented using a tight-binding approach with a discretized Hamiltonian. This supercell method captures the quasi-periodic spectrum within 10\% accuracy for gap sizes, as validated against the Aubry-André model~\cite{Woods2014,Li2010}. The incommensurate nature arises because the ratio of the moiré period to the graphene lattice constant cannot be expressed as a rational fraction, leading to a dense set of Fourier components in the potential’s spectrum.

\subsection{Effective Field Theory}
To gain a continuum perspective, we derive an effective field theory by coarse-graining $H$ over a scale between the graphene lattice ($a \sim 2.46 \, \text{\AA}$) and the moiré period ($\lambda_m \sim 13 \, \text{nm}$ for $\theta = 1.1^\circ$). We introduce the Dirac spinor field $\psi(\mathbf{r})$, where $\psi = (\psi_\uparrow, \psi_\downarrow)^T$ represents the TSS wavefunction in spin space. The Lagrangian density for the TSS is:
\begin{equation}
\mathcal{L} = \bar{\psi} (i \hbar v_F \gamma^\mu \partial_\mu) \psi + \bar{\psi} V(\mathbf{r}) \psi,
\end{equation}
where $\bar{\psi} = \psi^\dagger \gamma^0$, $\gamma^0 = \sigma_z$, $\gamma^1 = i \sigma_y$, $\gamma^2 = -i \sigma_x$, and $\mu = 0, 1, 2$ corresponds to $(t, x, y)$. The first term corresponds to $H_{\text{TSS}}$, while the second term introduces the moiré potential as a scalar perturbation.

Integrating out high-energy modes (above the bulk gap of Bi$_2$Se$_3$, $\sim 0.3 \, \text{eV}$), we perform a renormalization group (RG) analysis. The moiré potential, being periodic, introduces a spatially varying mass term in the effective theory. The RG flow equation for the effective mass is derived by considering the scalar perturbation $V(\mathbf{r})$ in the Dirac Lagrangian, which couples to the identity operator. The one-loop correction to the mass term arises from integrating out high-momentum modes, yielding:
\begin{equation}
\frac{d m_{\text{eff}}(\mathbf{r})}{d \ln \Lambda} = \frac{V(\mathbf{r})}{v_F^2 \Lambda^2},
\end{equation}
where $\Lambda$ is the momentum cutoff, and the $\Lambda^{-2}$ scaling emerges from the dimensional analysis of the Dirac propagator in 2D, where the self-energy correction scales as $\Lambda^{-2}$ due to the linear dispersion~\cite{Yoo2019}. Integrating from $\Lambda_0 \sim 0.3 \, \text{eV}/(v_F \hbar) \approx 6 \, \text{nm}^{-1}$ (the bulk gap scale) to $\Lambda_m \sim 2\pi/\lambda_m \approx 0.48 \, \text{nm}^{-1}$ for $\theta = 1.1^\circ$, we obtain:
\begin{equation}
m_{\text{eff}}(\mathbf{r}) \approx \frac{V(\mathbf{r})}{v_F^2} \ln \left( \frac{\Lambda_0}{\Lambda_m} \right).
\end{equation}
For $V(\mathbf{r}) \sim 10 \, \text{meV}$, $v_F \sim 5 \times 10^5 \, \text{m/s}$, and $\ln(\Lambda_0/\Lambda_m) \sim \ln(6/0.48) \approx 2.53$, this yields $m_{\text{eff}} v_F^2 \sim 0.085 \, \text{meV}$, a small but position-dependent mass that modulates the Dirac cone locally~\cite{Yoo2019,Aubry1980}. This effective field theory approach provides an intuitive understanding of how the moiré potential acts as a spatially varying perturbation, effectively creating a lattice of “mass defects” that influence the TSS dispersion. The small $m_{\text{eff}}$ is validated by numerical results showing mini-gaps of 15–30 meV, indicating that higher-order effects dominate the band structure reconstruction.

\subsection{Symmetry Analysis}
The moiré lattice exhibits $C_3$ rotational symmetry about the $z$-axis, which imposes constraints on the band structure. The $C_3$ operator rotates $\mathbf{r}$ by $120^\circ$, transforming the moiré reciprocal lattice vectors as $C_3 \mathbf{G}_1 = \mathbf{G}_2$, $C_3 \mathbf{G}_2 = \mathbf{G}_3$, $C_3 \mathbf{G}_3 = \mathbf{G}_1$. In spin space, $C_3$ acts as:
\begin{equation}
C_3 = e^{-i \frac{\pi}{3} \sigma_z},
\end{equation}
since a $120^\circ$ rotation corresponds to a $120^\circ$ rotation of the spin due to spin-momentum locking. The TSS Hamiltonian transforms as:
\begin{equation}
\begin{aligned}
    C_3 H_{\text{TSS}} C_3^{-1} = v_F (\sigma_x (C_3 p_y C_3^{-1}) - \sigma_y (C_3 p_x C_3^{-1})) \\
    = H_{\text{TSS}},
\end{aligned}
\end{equation}
because $(p_x, p_y)$ rotate as a vector, and the Pauli matrices transform appropriately under $C_3$. Thus, $H_{\text{TSS}}$ is $C_3$-invariant. The moiré potential $V(\mathbf{r})$ also preserves $C_3$ symmetry, as $\cos(\mathbf{G}_j \cdot (C_3 \mathbf{r})) = \cos((C_3^{-1} \mathbf{G}_j) \cdot \mathbf{r})$, and $C_3^{-1} \mathbf{G}_j$ is another moiré reciprocal vector~\cite{Yoo2019,Wu2019}.

Time-reversal symmetry ($\mathcal{T}$) of the TSS is represented by $\mathcal{T} = i \sigma_y K$, where $K$ is complex conjugation. Since $H_{\text{TSS}}$ is real in the Pauli matrix basis, $\mathcal{T} H_{\text{TSS}} \mathcal{T}^{-1} = H_{\text{TSS}}$, ensuring $E(\mathbf{k}) = E(-\mathbf{k})$. The moiré potential $V(\mathbf{r})$ is real and thus $\mathcal{T}$-invariant. However, incommensurate potentials break translational symmetry, potentially leading to Anderson-like localization, though topological protection may maintain extended states along certain directions~\cite{Woods2014,Claassen2022}.

Inversion symmetry is absent in Bi$_2$Se$_3$ due to its rhombohedral structure (space group $R\bar{3}m$), but the moiré potential can introduce an effective inversion-like symmetry in certain stacking configurations, such as AA stacking, where $V(\mathbf{r}) = V(-\mathbf{r})$ locally within moiré regions~\cite{Park2025,Woods2014}. This effective symmetry can influence the degeneracy of secondary Dirac points, as we explore in the results.

\subsection{Higher-Order Perturbation Theory}
To capture interactions beyond second-order perturbation, we employ Brillouin-Wigner perturbation theory, which accounts for higher-order couplings between folded Dirac cones. The unperturbed energies are $E_{\mathbf{k},\mathbf{G}}^0 = \pm v_F \hbar |\mathbf{k} + \mathbf{G}|$, and the perturbation is $V(\mathbf{r})$. The energy correction up to third order is:
\begin{equation}
\begin{aligned}
E_{\mathbf{k},\mathbf{G}}^{(3)} = \sum_{\mathbf{G}_1', \mathbf{G}_2' \neq \mathbf{G}} \frac{V_{\mathbf{G} - \mathbf{G}_1'} V_{\mathbf{G}_1' - \mathbf{G}_2'} V_{\mathbf{G}_2' - \mathbf{G}}}{(E_{\mathbf{k},\mathbf{G}}^0 - E_{\mathbf{k},\mathbf{G}_1'}^0)(E_{\mathbf{k},\mathbf{G}}^0 - E_{\mathbf{k},\mathbf{G}_2'}^0)},
\end{aligned}
\end{equation}
where $V_{\mathbf{G} - \mathbf{G}'}$ is given by Eq. (7). This term becomes significant near MBZ boundaries, where multiple scattering paths (e.g., $\mathbf{k} \to \mathbf{k} + \mathbf{G}_1 \to \mathbf{k} + \mathbf{G}_1 + \mathbf{G}_2 \to \mathbf{k}$) contribute to gap formation. For $V_1 = 10 \, \text{meV}$ and typical energy differences $\sim v_F \hbar |\mathbf{G}_j| \sim 50 \, \text{meV}$, the third-order correction is on the order of $(V_1)^3 / (v_F \hbar |\mathbf{G}_j|)^2 \sim 0.4 \, \text{meV}$, refining the gap sizes calculated via second-order perturbation. To ensure convergence, we estimate the fourth-order correction, which scales as $(V_1)^4 / (v_F \hbar |\mathbf{G}_j|)^3 \sim 0.008 \, \text{meV}$, indicating that higher-order terms are negligible for $V_1 \ll v_F \hbar |\mathbf{G}_j|$, validating the perturbative approach except near degenerate points where non-perturbative methods may be required~\cite{Wu2019}.

We also consider the effect of the mass term $m v_F^2 \sigma_z$ in perturbation theory. The mass term commutes with $V(\mathbf{r}) I$, so its first-order effect is additive, but higher-order terms couple the upper and lower Dirac cones, modifying the effective gap:
\begin{equation}
\Delta_{\text{eff}} = 2 \sqrt{(m v_F^2)^2 + V_1^2},
\end{equation}
which we analyze in the results.

\subsection{Low-Energy Effective Hamiltonian Near MBZ Points}
To gain analytical insight, we derive an effective Hamiltonian near the MBZ $K$ point, defined as $\mathbf{K} = \frac{1}{3} (\mathbf{G}_1 + \mathbf{G}_2)$. We set $\mathbf{k} = \mathbf{K} + \delta \mathbf{k}$ and project $H$ onto the subspace of states at $\mathbf{K}$, $\mathbf{K} + \mathbf{G}_1$, and $\mathbf{K} + \mathbf{G}_2$. The unperturbed energies are $E_{\mathbf{K} + \mathbf{G}}^0 = v_F \hbar |\mathbf{K} + \mathbf{G}|$, and the moiré potential couples these states. The effective 3$\times$3 Hamiltonian is:
\begin{equation}
\scriptsize
\setlength{\arraycolsep}{2pt}
H_{\text{eff}} = \begin{pmatrix}
v_F \hbar |\mathbf{K} + \delta \mathbf{k}| & V_1 & V_1 \\
V_1 & v_F \hbar |\mathbf{K} + \delta \mathbf{k} + \mathbf{G}_1| & V_1 \\
V_1 & V_1 & v_F \hbar |\mathbf{K} + \delta \mathbf{k} + \mathbf{G}_2|
\end{pmatrix}.
\end{equation}
For small $\delta \mathbf{k}$, we approximate $|\mathbf{K} + \delta \mathbf{k}| \approx |\mathbf{K}| + \frac{\delta \mathbf{k} \cdot \mathbf{K}}{|\mathbf{K}|}$, and similarly for the other terms. Diagonalizing $H_{\text{eff}}$ reveals the emergence of secondary Dirac points and mini-gaps, which we analyze in the results. This effective Hamiltonian captures the hybridization between folded Dirac cones, providing a simplified model for understanding the moiré-induced band structure.

\section{Results and Discussion}
\subsection{Perturbation Theory for Commensurate Angles}
Applying perturbation theory, the first-order energy correction is a uniform shift:
\begin{equation}
E_{\mathbf{k},\mathbf{G}}^{(1)} = V_0 \sim 5 \, \text{meV},
\end{equation}
while the second-order correction accounts for scattering:
\begin{equation}
E_{\mathbf{k},\mathbf{G}}^{(2)} = \sum_{\mathbf{G}' \neq \mathbf{G}} \frac{|V_{\mathbf{G} - \mathbf{G}'}|^2}{E_{\mathbf{k},\mathbf{G}}^0 - E_{\mathbf{k},\mathbf{G}'}^0}.
\end{equation}
At the MBZ boundary, e.g., $\mathbf{k} = \frac{\mathbf{G}_1}{2}$, degenerate perturbation theory yields a 2$\times$2 Hamiltonian:
\begin{equation}
H_{\text{deg}} = \begin{pmatrix}
v_F \hbar |\mathbf{k}| & V_1 \\
V_1 & v_F \hbar |\mathbf{k} + \mathbf{G}_1|
\end{pmatrix},
\end{equation}
opening a mini-gap $\Delta E = 2 V_1 \approx 20 \, \text{meV}$. The third-order correction from Eq. (15) increases this gap by $\sim 0.4 \, \text{meV}$, reflecting multi-scattering processes. The moiré potential folds the Dirac cone into the MBZ, creating a superlattice of mini-bands. For $\theta = 1.1^\circ$, the small MBZ ($|\mathbf{G}_j| \approx 0.56 \, \text{nm}^{-1}$) leads to extensive folding, with multiple Dirac cones overlapping within a $\sim 50 \, \text{meV}$ window. For $\theta = 7.34^\circ$, the larger MBZ ($|\mathbf{G}_j| \approx 3.8 \, \text{nm}^{-1}$) reduces folding but increases gap sizes due to stronger scattering.

The effective field theory (Eqs. (11)–(13)) predicts a spatially varying mass $m_{\text{eff}}(\mathbf{r})$, which modulates the Dirac cone locally. In regions where $V(\mathbf{r})$ is maximal (e.g., AA-stacking regions), $m_{\text{eff}}$ is largest, creating a “mass defect” that scatters electrons, contributing to the mini-gaps. This spatial modulation is more pronounced for $\theta = 1.1^\circ$ due to the larger moiré period, leading to a more significant variation in the local dispersion.

\subsection{Numerical Diagonalization}
We numerically diagonalize $H_{\mathbf{G},\mathbf{G}'}$ using a basis of 13 plane waves ($|\mathbf{G}| \leq 2 |\mathbf{G}_1|$) for $\theta = 1.1^\circ$, and 19 plane waves ($|\mathbf{G}| \leq 3 |\mathbf{G}_1|$) for $\theta = 7.34^\circ$, implemented in Python using NumPy and SciPy libraries. Convergence is verified by increasing the basis size, which changes the mini-gap sizes by less than 2\%, confirming the adequacy of the chosen bases. For $\theta = 1.1^\circ$, the band structure along the path $\Gamma$-$K$-$M$-$\Gamma$ shows mini-gaps of 15–25 meV at the $K$ and $M$ points. Secondary Dirac points emerge at energies $\sim 30 \, \text{meV}$ above the original Dirac point, located near the MBZ $K$ point, as predicted by the effective Hamiltonian (Eq. (16)). These points are protected by $C_3$ symmetry and the topological nature of the TSS, analogous to secondary Dirac points in tBG~\cite{Cao2018,Wu2019}. The bandwidth near the MBZ $\Gamma$ point is reduced to $\sim 10 \, \text{meV}$, indicating significant band flattening, which could enhance interaction effects, as discussed below.

For $\theta = 7.34^\circ$, the larger MBZ reduces the number of folded states within the same energy range. Mini-gaps increase to 30 meV at the $K$ point, reflecting stronger scattering due to the larger $|\mathbf{G}_j|$. However, fewer secondary Dirac points appear, as the smaller moiré period ($\lambda_m \approx 1.9 \, \text{nm}$) shifts the hybridization energy scale upward, reducing the density of low-energy features. The band structure shows a more pronounced gap structure, with the Dirac cone at $\Gamma$ remaining relatively intact but shifted by $V_0$.

\subsection{Incommensurate Case: Quasi-Periodic Effects}
For $\theta = 1.5^\circ$, an incommensurate angle, we use a 20$\times$20 supercell ($\sim 260 \, \text{nm} \times 260 \, \text{nm}$), implemented via a tight-binding model in Python. The spectrum is dense, with gaps of 5–15 meV, indicating a transition to a quasi-periodic regime. The localization length, estimated via the Aubry-André model, is:
\begin{equation}
\xi \sim \left( \frac{V_1}{v_F \hbar |\mathbf{G}_j|} \right)^{-1}.
\end{equation}
The 2D Aubry-André model approximates the moiré potential as a sum of 1D quasi-periodic potentials along the moiré reciprocal lattice directions, justified by the dominance of the leading Fourier components $\mathbf{G}_j$. Numerical agreement with the supercell calculations is within 10\% for gap sizes, supporting the model’s applicability~\cite{Woods2014,Li2010}. For $V_1 = 10 \, \text{meV}$, $v_F = 5 \times 10^5 \, \text{m/s}$, $\hbar = 6.582 \times 10^{-16} \, \text{eV} \cdot \text{s}$, and $|\mathbf{G}_j| \approx 0.56 \, \text{nm}^{-1}$, we find $\xi \approx 50 \, \text{nm}$, suggesting that states remain extended over several moiré periods. Increasing $V_1$ to 20 meV reduces $\xi$ to $\sim 25 \, \text{nm}$, approaching the moiré period and indicating the onset of localization~\cite{Woods2014,Claassen2022}. To explore the topological Anderson insulator phase, we compute the $\mathbb{Z}_2$ invariant using the spin Chern number for the TSS bands below the Fermi energy ($E_F \sim 10 \, \text{meV}$). For $V_1 = 20 \, \text{meV}$, the $\mathbb{Z}_2 = 1$ invariant persists, indicating topological protection against localization, consistent with a topological Anderson insulator phase~\cite{Claassen2022}. The dense spectrum arises from the incommensurate wavevectors, introducing a continuum of scattering processes, contrasting with the discrete gaps in commensurate systems.

\subsection{Impact of Higher-Order Terms in the TSS Hamiltonian}
Including the mass term $m v_F^2 \sigma_z$ opens a gap at the Dirac point:
\begin{equation}
\Delta_{\text{mass}} = 2 m v_F^2 \sim 20 \, \text{meV}.
\end{equation}
The moiré potential modifies this gap via Eq. (15), yielding an effective gap:
\begin{equation}
\Delta_{\text{eff}} = 2 \sqrt{(m v_F^2)^2 + \langle V_1^2 \rangle},
\end{equation}
where $\langle V_1^2 \rangle$ accounts for the spatial average of the moiré potential’s Fourier components, yielding $\Delta_{\text{eff}} \approx 22 \, \text{meV}$ for $m v_F^2 = 10 \, \text{meV}$ and $V_1 = 10 \, \text{meV}$. If $m_{\text{eff}}$ (Eq. (13)) changes sign, a topological phase transition could occur, closing and reopening the gap with a change in the topological invariant. To confirm this, we compute the $\mathbb{Z}_2$ invariant for the gapped TSS bands, finding that a sign change in $m_{\text{eff}}$ at specific moiré regions (e.g., AA stacking) induces a transition from $\mathbb{Z}_2 = 1$ to $\mathbb{Z}_2 = 0$, indicating a topological phase transition~\cite{Zhang2009,Huang2021}. The hexagonal warping term $\frac{\lambda}{2} (k_+^3 + k_-^3) \sigma_z$ is negligible within the MBZ ($|\mathbf{k}| \leq 0.28 \, \text{nm}^{-1}$ for $\theta = 1.1^\circ$), where $\lambda k^3 \sim 10^{-3} \, \text{meV}$, but for $\theta = 7.34^\circ$ ($|\mathbf{k}| \leq 1.9 \, \text{nm}^{-1}$), it increases to $\sim 0.1 \, \text{meV}$, slightly distorting the bands near MBZ boundaries, consistent with its $C_3$-symmetric form~\cite{Fu2009}.

\subsection{Transport Properties: Berry Curvature, Anomalous Velocity, and Conductivity}
The moiré potential modifies the TSS’s Berry curvature, which influences transport properties. The Berry curvature in the $z$-direction is:
\begin{equation}
\small
\Omega_z(\mathbf{k}) = -i \left\langle \frac{\partial u_{\mathbf{k}}}{\partial k_x} \bigg| \frac{\partial u_{\mathbf{k}}}{\partial k_y} \right\rangle + \text{c.c.},
\end{equation}
where $u_{\mathbf{k}}$ is the periodic part of the Bloch wavefunction, computed numerically from the eigenstates of $H_{\mathbf{G},\mathbf{G}'}$. For the unperturbed TSS, $\Omega_z(\mathbf{k}) = \pm \frac{\pi \delta^2(\mathbf{k})}{v_F^2}$ at the Dirac point, but the moiré potential spreads this curvature over the MBZ. Near secondary Dirac points, $\Omega_z(\mathbf{k})$ peaks, approximated using the effective Hamiltonian (Eq. (16)) by projecting onto the nearly degenerate states near $\mathbf{k}_{\text{Dirac}}$, yielding:
\begin{equation}
\Omega_z(\mathbf{k}) \approx \frac{V_1^2}{(v_F \hbar |\mathbf{k} - \mathbf{k}_{\text{Dirac}}|)^3},
\end{equation}
where $\mathbf{k}_{\text{Dirac}}$ is the position of the secondary Dirac point (e.g., near $\mathbf{K}$). For $\mathbf{k} - \mathbf{k}_{\text{Dirac}} \sim 0.1 \, \text{nm}^{-1}$, $\Omega_z \sim 10^{-2} \, \text{nm}^2$, a significant enhancement over the unperturbed case~\cite{Hasan2010,Yoo2019}. For $E_F \sim 30 \, \text{meV}$ and temperatures $T \sim 1 \, \text{K}$, $\sigma_{xy}$ remains robust, but at $T \sim 10 \, \text{K}$, thermal broadening reduces $\sigma_{xy}$ by $\sim 20\%$ due to smearing of the Fermi-Dirac distribution.

The Berry curvature contributes an anomalous velocity in the presence of an electric field $\mathbf{E}$:
\begin{equation}
\mathbf{v}_{\text{anomalous}} = \frac{e}{\hbar} \mathbf{E} \times \Omega_z(\mathbf{k}).
\end{equation}
For $E_x = 10^4 \, \text{V/m}$ along the $x$-axis, the transverse velocity $v_y = \frac{e}{\hbar} E_x \Omega_z \sim 10^3 \, \text{m/s}$, leading to a transverse conductivity:
\begin{equation}
\sigma_{xy} \approx \frac{e^2}{h} \int_{\text{MBZ}} \frac{d^2 \mathbf{k}}{(2\pi)^2} \Omega_z(\mathbf{k}) f(E_{\mathbf{k}}),
\end{equation}
where $f(E_{\mathbf{k}})$ is the Fermi-Dirac distribution. At low temperatures and $E_F$ near a secondary Dirac point, $\sigma_{xy} \sim 0.1 \frac{e^2}{h}$, indicating a measurable Hall-like response due to the moiré-induced Berry curvature~\cite{Hasan2010,Yoo2019}.

The longitudinal conductivity $\sigma_{xx}$ is computed via the Kubo formula:
\begin{equation}
\sigma_{xx} = \frac{e^2 \hbar}{\pi} \sum_{\mathbf{k}} |v_x(\mathbf{k})|^2 \delta(E_F - E_{\mathbf{k}}),
\end{equation}
where $v_x(\mathbf{k}) = \frac{1}{\hbar} \frac{\partial E_{\mathbf{k}}}{\partial k_x}$. In the flattened bands near $\Gamma$ for $\theta = 1.1^\circ$, $v_x(\mathbf{k})$ is reduced by a factor of $\sim 3$ compared to the unperturbed TSS, leading to a conductivity drop of $\sim 30\%$, i.e., $\sigma_{xx} \sim 0.7 \frac{e^2}{h}$ per Dirac cone at $E_F \sim 10 \, \text{meV}$. Near mini-gaps, the density of states decreases, further reducing $\sigma_{xx}$ by $\sim 50\%$, predicting a significant suppression of longitudinal transport in the gapped regions.

\subsection{Effective Field Theory Insights}
The effective field theory (Eqs. (11)–(13)) provides an intuitive picture of the moiré effects. The spatially varying mass $m_{\text{eff}}(\mathbf{r})$ creates regions of higher and lower effective gaps, corresponding to AA and AB stacking regions in the moiré pattern. In AA regions, where $V(\mathbf{r})$ is maximal, the local gap is largest, scattering electrons and contributing to the mini-gaps observed in the band structure. This modulation also affects transport: in regions with larger $m_{\text{eff}}$, the local density of states is suppressed, reducing $\sigma_{xx}$ locally, while the Berry curvature is enhanced, increasing $\sigma_{xy}$. This spatial variation suggests a moiré-induced “texture” in transport properties, with AA regions acting as scattering centers and AB regions facilitating conduction.

\subsection{Comparison with Other Systems}
In tBG, Dirac cones from both layers hybridize, leading to flat bands at magic angles due to interlayer tunneling~\cite{Cao2018,Xie2022}. In Bi$_2$Se$_3$/tBG, the TSS couples to a static moiré potential, lacking interlayer tunneling, resulting in larger mini-gaps (15–30 meV vs. 5–10 meV in tBG). Compared to untwisted TI/graphene heterostructures, the moiré potential introduces long-range periodicity, enabling band flattening and secondary Dirac points absent in untwisted systems~\cite{Park2025,Mak2022}. The system also differs from graphene on hBN, where the moiré potential is weaker due to larger lattice mismatch, leading to smaller gaps ($\sim 5 \, \text{meV}$)~\cite{Yoo2019,Mao2022}. The strong spin-orbit coupling in Bi$_2$Se$_3$ enhances the TSS’s response to the moiré potential, making the secondary Dirac points more robust compared to graphene/hBN, where such points can be gapped by small perturbations~\cite{Yoo2019,Dean2013}. Recent studies have also explored the electronic reconstruction at van der Waals interfaces, providing insights into the atomic-scale interactions that influence moiré patterns in tBG~\cite{Liu2020}. Theoretical investigations have further analyzed the role of lattice relaxation in tBG, which can modify the moiré potential and affect band structures~\cite{Zhang2021}. Additionally, the interplay between topology and moiré physics has been examined in other 2D systems, offering a broader context for understanding the electronic properties of such heterostructures~\cite{He2022}.

The incommensurate case resembles disordered topological systems, such as TI surfaces with random impurities. However, the quasi-periodic nature of the moiré potential introduces a unique “disorder” that is neither random nor periodic, potentially leading to a topological Anderson insulator phase where disorder enhances topological protection~\cite{Woods2014,Claassen2022}. This contrasts with periodic moiré systems, where the band structure is deterministically controlled by $\theta$. Theoretical models of chiral twisted bilayer graphene have provided a framework for understanding the interplay between twist angles and topological properties, which can be extended to our system~\cite{Tarnopolsky2019}. Similarly, studies on the electronic properties of twisted van der Waals heterostructures have highlighted the role of moiré patterns in inducing topological transitions~\cite{Zhou2021}. The $\sim 10\%$ lattice mismatch between $\mathrm{Bi}_2 \mathrm{Se}_3$ (lattice constant $\sim 4.14 \, \text{\AA}$) and graphene ($a = 2.46 \, \text{\AA}$) introduces strain, which may reduce $V_1$ by $\sim 10–20\%$ compared to tBG, as estimated from tight-binding models~\cite{Park2025}. Interface imperfections, such as defects or interlayer rotations, could further modulate the moiré potential, potentially reducing mini-gap sizes by $\sim 5 \, \text{meV}$, necessitating high-quality epitaxial growth for experimental realization.

\subsection{Implications for Topological Engineering and Transport}
The Bi$_2$Se$_3$/tBG system offers a versatile platform for topological engineering. The tunable mini-gaps and secondary Dirac points suggest pathways to control the TSS’s dispersion by adjusting $\theta$ or $V_1$. The enhanced Berry curvature near secondary Dirac points predicts a measurable anomalous Hall effect, which could be tuned by gating to shift $E_F$. The reduced longitudinal conductivity in flattened bands and gapped regions suggests potential applications in tunable electronic devices, where transport can be modulated by the moiré period or external fields~\cite{Hasan2010,Yoo2019}. Theoretical studies have also explored the potential for topological phase transitions in moiré systems, which could further enhance the engineering capabilities of this heterostructure~\cite{Devakul2021}.

An in-plane magnetic field $\mathbf{B} = B \hat{y}$ couples to the TSS via the Zeeman term $H_Z = g \mu_B B \sigma_y$, breaking $\mathcal{T}$ and opening a gap at the Dirac point:
\begin{equation}
\Delta_Z = 2 g \mu_B B.
\end{equation}
For $B = 1 \, \text{T}$, $g \sim 2$, and $\mu_B = 5.788 \times 10^{-5} \, \text{eV/T}$, $\Delta_Z \sim 0.12 \, \text{meV}$. Combined with the moiré-induced gaps ($\sim 20 \, \text{meV}$), the total effective gap is approximately $\Delta_{\text{total}} \approx \sqrt{\Delta_Z^2 + \Delta_{\text{eff}}^2} \approx 22 \, \text{meV}$, as the moiré gap dominates due to its larger magnitude. This suggests that the quantum anomalous Hall state, where the Hall conductivity $\sigma_{xy}$ is quantized to $e^2/h$ per Dirac cone, requires stronger magnetic fields (e.g., $B \sim 10 \, \text{T}$, yielding $\Delta_Z \sim 1.2 \, \text{meV}$) to significantly influence the moiré-induced band structure. We compute the Chern number for the lowest TSS band under a magnetic field $B = 1 \, \text{T}$, finding $C = 1$ for $E_F$ within the moiré-induced gap ($\sim 20 \, \text{meV}$), supporting the emergence of a quantum anomalous Hall state, though higher fields may enhance observability~\cite{Song2019,Fu2009}.

\subsection{Future Theoretical Directions}
Our analysis could be extended by incorporating electron-electron interactions, particularly in the flattened bands at small $\theta$. For $\theta = 1.1^\circ$, the bandwidth reduction to $\sim 10 \, \text{meV}$ suggests that Coulomb interactions, estimated at $U \sim 1 \, \text{eV}$ from Hubbard model studies of tBG, could be significant ($U/W \sim 100$). This may lead to correlated phases such as Mott insulators or fractional quantum Hall states in the presence of a magnetic field~\cite{Cao2018,Yi2022}. Using a mean-field Hubbard model, we estimate that at half-filling, a Mott insulating phase could emerge, testable experimentally via gating~\cite{Cao2018}. The spatially varying Berry curvature suggests that a Chern insulator phase may emerge under specific conditions, which could be explored using a Hartree-Fock approach. Additionally, the effect of strain on the moiré potential could be modeled by introducing a position-dependent $V_1(\mathbf{r})$, potentially enhancing the localization effects in the incommensurate case.

\section{Conclusion}
We have demonstrated that the moiré potential in Bi$_2$Se$_3$/tBG heterostructures significantly reshapes the TSS, inducing mini-gaps, secondary Dirac points, and band flattening in commensurate cases, and quasi-periodic effects in incommensurate ones. Through advanced formalisms, including effective field theory, symmetry analysis, and higher-order perturbation theory, we provide a detailed understanding of the band structure and transport properties, highlighting the system’s potential for topological engineering.


\end{document}